*BAHI Camile, 2018*

*MPharm*

**ᴜNIVERSITĕ** ᴅᴇ
**FRANCHE-COMTĕ**

# Psilocybin based therapy for cancer related distress, a systematic review and meta analysis

# Summary

# Introduction

Depression and anxiety are pretty common in patients with cancer, 30 to 40 % (Mitchell et al. 2011) of the patients express these symptoms, which are often linked with several problems like a decrease of quality life, non adherence to medical treatments, increased healthcare utilization, decrease of social function, increased sensitivity to pain, hopelessness, desire of dying, decreased survival rates, increased rates of suicide. (Arietta et al. 2014)

Antidepressant and benzodiazepine (less frequently) are usually used to treat these symptoms but the efficiency of these medicines is limited. (Ostuzzi et al. 2015) Benzodiazepine are mainly designed for a short term use, and have a negative benefices/risks balance for a long term range treatment, involving addiction, tolerance, higher risk of overdosing, loss of memory and social function; (Ashton 1986). Moreover, antidepressants never showed its superiority to placebo for treating cancer related distress (Ostuzzi et al., 2015)

A serotoninergic psychedelic substance, psilocybin (psilocyn) that is 5-HT2A receptors agonist (Vollenweider and Kometer, 2010) and induces various and importants modifications in perceptions, emotions and sensations (Halberstadt 2015) appear to be promising for treating cancer related depression and anxiety in various recent studies. Serotoninergic psychedelics has been tested for this purpose back in the 60s and 70s and the resulting studies suggested that theses molecules might be useful in treating distress for cancer patients, (Grof et al. 1973) but these studies probably does not stick to the actual research standards.

Unfortunately, the research about serotonin agonists psychedelics have been stopped  for many years in response to the wide-spread non medical use of these substances, until a few years ago. Which also caused a bad reputation towards these substances.

Very recently, some research institutes have taken back psychedelics research, and proved that this molecule has a well-established physiological and psychological safety profile in human laboratory and clinical trial research (Johnson et al. 2008), is not known to be addictive and may have anti-addictive properties (Ross, 2012).

Heffter institute is the main organism leading studies about these substances and trials for treatment against anxiety and depression (mainly cancer related depression), and addiction.

This review aims to examine if a psilocybin based therapy could be considered for treating patients with cancer related depression and/or anxiety and more precisely if it can be considered as efficient and also safe for this purpose with the actual knowledge.

It also aims to sum up the work Heffter institute have accomplished until now on this particular subject, and eventually to get a better understanding about psychological and neurobiological mechanisms implicated in the anti depressant and anxiolytics action of psilocybin, if it is confirmed.

# Method

## Realization of the review

This systematic review is inspired by the PRISMA directives lines, according to the ISSM PRISMA checklist that can be found on https://www.elsevier.com/__data/promis_misc/ISSM_PRISMA_Checklist.pdf
Ethical approval for this protocol and the planned systematic review was not required.

## Searching process

Since this review focuses on Heffter institute research works, the research procedure consisted in reaching the organism database in the « cancer » section and looking for the studies available which can be found on : https://heffter.org/study-publications/. Further research have been reached on PubMed to august 2018, but no other relevant study met the inclusion criteria, with the keywords : psilocybin, depression, anxiety, cancer (together).

## Selecting the studies

The search returned 15 articles on Pubmed and 5 on the Heffter institute database. On the 20 articles, 4 met the inclusion criteria and have been analyzed in this systematic review.

**Inclusion criteria included :**

– Being about the use of psilocybin for treating cancer related depressive and anxious states
– Including adult participants (18+ years old) with SCID (DSM-IV) diagnosis Adjustment disorder w/anxiety and/or depressed mood, chronic
– Being published in English language at the date of the 16 august 2018.
– Including at least 20 participants for the quantitative studies, 10 for the qualitative ones
– Being randomized, blinded if quantitative study

Studies were excluded if they were not published in an English language peer-reviewed journal, or if they used others substances than psilocybin to treat depression and anxiety.

several comparisons of the studies have been done  to determine if  some were too similar in order to exclude those from the review. In the present case, none of these studies has been excluded.
Data have been extracted after multiple readings through piloted forms for quantitative studies and independently for qualitative studies.

## Data extraction

Extracted data included : author, year of publication, country, study design, sample size, percentage of women, type of cancer, stage, follow up, studied parameters, measurement tools, prior psychiatric conditions, Adverse events, Study design, Dosages, number of dosage sessions, occurrence of serious psychiatric adverse events and quantitative data for analysis.

Data has been manually extracted trough pre filled form using «apple's Numbers » application for Mac OS.

## Quality assessment

The overall quality of the studies and the risk of biais has been assessed according Cochrane Risk of Bias Tool for Randomized Controlled Trials (Cochrane collaboration, 2008) and the Jadad scale (Hempel et al. 2011) when it came to quantitative studies and to Critical Appraisal Skills Programme for qualitative studies when it came to qualitative studies (Critical Appraisal Skills Program, 2018 ), these tool provides guidance in order to evaluate the internal validity of the study and for translating this to risk of bias. Each study has been rated as being as good quality, fair quality and poor quality.

Performing a funnel plot analysis was not relevant according to the relatively restricted amount of studies.

## Results synthesis and outcomes :

Data about depressed patients's response to psilocybin, superiority from placebo at different periods of follow up on depression and anxiety,  psychiatric and medical safety, psychological and neurobiological supposed mechanisms, has been presented through a systematic narrative synthesis.

*BAHI Camile, 2018*

A peto Odds ratio (statsdirect 2018) has been manually performed for a meta analysis and integrated into forest plots to compare the quantitative results exploited in the studies. These analyses were synthetized in separate tables.

# Results

## Included studies

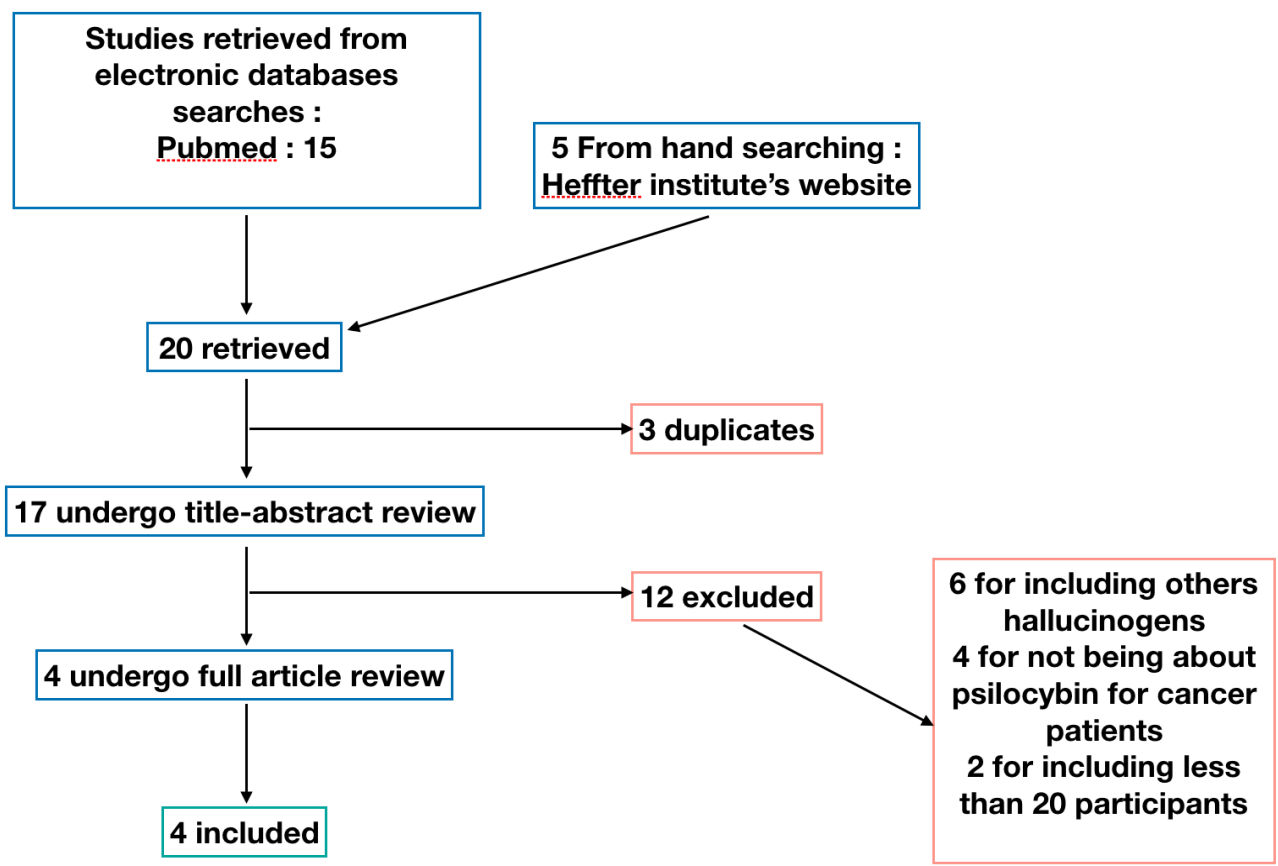

**Fig 1 : PRISMA (Preferred Reporting Items for Systematic reviews and Meta-analyses) diagram**

20 articles were screened for this review.

After reading the papers, excluding duplicates, 4 met the inclusion criteria, (see Figure 1 for flowchart).

2 of these articles (Ross et al. 2016, Griffiths et al. 2016) consisted in randomized double blind crossover quantitative studies with cancer patients, in one unique do-sage session of psilocybin as treatment for cancer related distress with a 6-month follow-up.

Both used a medium to high dose of psilocybin (0,3mg/kg and 30mg/70kg) and compared its effect to a placebo, niacin in the study conducted by Ross and colleagues, and a low dose of psilocybin (1mg/kg), considered as inactive, in the study conducted by Griffiths and colleagues.

A total of 79 patient received psilocybin in these two studies.

2 studies (Belser et al. 2017, Swift et al. 2017) consisted in 13 qualitative interviews each about understanding the psychological mechanisms involved in a psilocybin session and getting a better understanding about it's anti depressive and anxiolytic effect on cancer patients (interview were realized with participants coming from the first quantitative study, conducted by Ross and colleagues and included in this review), 5 of the participants were interviewed within 1 week following their second psilocybin dosing session, and eight were interviewed approximately 1 year following their session.

## Characteristics of the available literature

The extracted studies characteristics were presented in supplementary table 1 (available page 31 and in a numeric format on demand).

Of the two quantitative studies, both studied similar parameters (systolic and diastolic blood pressure, hearth rate, depression, anxiety, subjective drug effect, quality of life, sustaining of the effects, spirituality, spiritual well being)

Similar scales were also shared between these studies to analyze and quantify depression and anxiety such as : Hospital Anxiety and Depression Scale (HADS), self-rated subscale of depression (HAD-D) (Snaith 2003),  Beck Depression Inventory (BDI) (Beck 1961); Spielberger State-Trait Anxiety Inventory (STAI) (american psychological association 2018) self-report measure of state (STAI state or STAI S) and trait (STAI trait or STAI T) anxiety… (Griffiths et al. 2016, Ross et al.2016)

The second study, conducted by Griffiths and colleagues used several more scales to quantify some more precise secondary outcomes, as mystical experience for example, for further details, please refer to supplementary table 1.

The two qualitative studies focused on getting a better comprehension of the psychological mechanisms underlying the anxiolytic and anti-depressant effects of psilocybin.

For this purpose, interviews were piloted by an interpretative Phenomenological Analysis (IPA) (Smith and Osborn 2007), consisting in "a qualitative research approach committed to the examination of how people make sense of their major life experiences". (Belser et al. 2017, Swift et al. 2017)

This questionnaire included question about meaningful visuals, emotions, wisdom lessons, psychological distress, feeling of interconnectedness, letting go, shift in perspectives toward cancer recurrence or progression… (Belser et al 2017. Swift et al. 2017)

## Participant and clinical features of samples in included studies

The 4 studies included a total of 105 randomized participants, with the number of randomized participants varying from 13 to 51, there was slightly more women overall (66%). The mean age across the studies was approximately 56 and the ages within the studies, when it was available, varied from 22 to 75.

Psychiatric conditions were analyzed thanks to DSM-IV (Bell, 1994) in all studies, and before the first session, a total of 76 participants presented an SCID (DSM-IV) diagnosis Adjustment disorder with anxiety and depressed mood, chronic Or SCID (DSM-IV) diagnosis Adjustment disorder with anxiety, chronic. 5 participants presented a diagnosis of dysthymic disorder, 8 with generalized anxiety disorder (GAD), 14 with major depressive disorder (MDD) , 4 participants presented a dual diagnosis of GAD with MDD and 1 participant was diagnosed a GAD and dysthymic disorder. (Ross et al. 2016, Griffiths et al. 2016, Belser et al. 2017, Swift et al. 2017)

## Medication Status

In both quantitative studies, none of the participants were under any psychotropic substance at the time of the studies enrollment.

Approximately 55% of the participants had been under anti-depressant or anxiolytic medication prior the overall studies. (Ross et al 2016. Griffiths et al. 2016)

## Risk of Biais within the included studies

Quantitative Studies were appraised using Cochrane Risk of Bias Tool for Randomized Controlled Trials and Jadad scale, complete results are available in table 1 and 2.

Both studies obtained 4 out of 5 points at the Jadad scale which is qualified as rigorous and filled all items of the Cochrane's checklist (Cochrane collaboration , 2008), but 1 uncertain concerning the blinding method for the study conducted by Ross and colleagues.

According to the scores obtained using these 2 tools , both studies obtain a « good quality » ranking and appear to have a low risk of biais.

Qualitative studies where appraised using Critical Appraisal Skills Program for qualitative studies (CASP, 2018) and met all criteria. According to this tool, the studies are ranked as « good quality studies » with a low risk of biais (more detail in table 3).

**Table 1. Cochrane Risk of Bias Tool for Randomized Controlled Trials**

| | Random sequence generation | Allocation concealment | Selective reporting | Others biais | Blinding of participants and personnel | Blinding of outcome assessment | Incomplete outcome data | Overall quality of the study |
|---|---|---|---|---|---|---|---|---|
| **Study 1 : Ross et al.** | Low risk of biais | Low risk Of biais. (Sequentially numbered drug containers of identical appearance) | Low risk of biais (The study protocol is available and all of the study's pre-specified (primary and secondary) outcomes that are of interest in the review have been reported in the pre-specified way) | None | Unclear risk of biais (not enough information) | Low risk of biais (explained in qualitative studies) | Low risk of biais (No missing outcome data) | Good quality (2 unclear risk of biais, others low) |
| **Study 2 : Griffiths et al.** | Low risk of biais | Low risk Of biais Sequentially numbered drug containers of identical appearance) | Low risk of biais (The study protocol is available and all of the study's pre-specified (primary and secondary) outcomes that are of interest in the review have been reported in the pre-specified way) | None | Low risk of biais ( Blinding of participants and key study personnel ensured, and unlikely that the blinding could have been broken) | Low risk of biais ( Blinding of outcome assessment ensured, and unlikely that the blinding could have been broken.) | Low risk of biais  (No missing outcome data ) | Good quality (all criteria filled low risk of biais) |

**Table 2. Jadad scale**

| | Was the study described as randomized ? | Was the study described as double blind ? | Was there a description of Withdrawals and dropouts ? | Optional points |
|---|---|---|---|---|
| **Study 1 : Ross et al.** | Yes | Yes | Yes | Method of blinding explained and ap-propriated |
| **Study 2 : Griffiths et al.** | Yes | Yes | Yes | Method of blinding explained and ap-propriated |

**Table 3. Critical Appraisal Skill Program for quality studies**

| CASP for quality reviews | Was there a clear statement of the aims of research ? | Is a qualitative methodology appropriate ? | Was the research design appropriate to address the aims of research ? | Was the recruitment strategy appropriate to the aims of the research ? | Was the data collected in a way that addressed the research issue | Has the relationship between researcher and participants ? | Have ethical issue been taken into consideration ? | Was the data analysis sufficiently rigorous ? | Is there a clear statement of the findings ? | How valuable is the research |
|---|---|---|---|---|---|---|---|---|---|---|
| **Study 3 : Belser et al.** | Yes (In the current analysis, we address research questions regarding the form and content of participant experiences during the psilocybin dosage sessions, descriptions of their subjective experiences of this psychological intervention in context, and their understandings of the embedded meanings of their lived experiences.) | Yes (concerning subjectives experiences) | Yes (well structured and explained interviews with precise IPA questionnaire) | Yes (follow up from phase 2 psilocybin clinical trial) | Yes (interviews in person but one audio conference, audio recorded and transcribed verbatim, methodology well explained (IPA)) | Yes (successive analysis of the collected data by different reviewer to avoid bias) | Yes (Participants were informed that agreeing to be interviewed was entirely optional and would have no effect on their involvement in the quantitative study. They were offered no financial compensation, and there were no penalties for declining to participate. All volunteers gave their informed consent prior to participation. The study was approved by the institutional review board of the NYU School of Medicine.) | Yes (in depth data analysis protocole available in the study, limitations of the interviews taken in consideration, data selection explained) | Yes (explicit discussed findings, discussion in relation to the original research question) | Highly valuable :(first qualitative study about that subject, based on a promising quantitative study, identifies new area where research can be necessary, useful to understand overall psilocybin psychological mechanisms) |
| **Study 4 : Swift et al.** | Yes (The current article explores psilocybin therapy experiences directly related to cancer and death, and together with the previous analysis aims to capture a more complete understanding of the treatment through patient narrative accounts) | Yes (concerning subjectives experiences) | Yes (well structured and explained interviews with precise IPA questionnaire) | Yes (follow up from phase 2 psilocybin clinical trial) | Yes ( interview by teleconference, transcribed verbatim, well explained methodology (IPA questionnaire)) | Yes (successive analysis of the collected data by different reviewer to avoid bias) | Yes : (All participants were informed that participation in this qualitative study was entirely optional, and their decision to enroll would not affect their involvement in the original study. Participants received no financial compensation, nor were there penalties for declining to participate. All participants provided written informed consent prior to enrolling. Pseudonyms are used throughout the article to protect participant confidentiality. All participants were still alive at the time of publication.) | Yes (in depth data analysis protocole available in the study, limitations of the interviews taken in consideration, data selection explained) | Yes (explicit discussed findings, discussion in relation to the original research question) | Highly valuable (giving a further point of view on different aspects after the first qualitative study about that subject, based on a promising quantitative study, identifies new areas where research can be necessary, useful to understand overall psilocybin psychological mechanisms) |

# Study Results

## Depression

The two quantitative studies evaluated psilocybin as potential treatment for depression in patients with cancer (Ross et al. 2016, Griffiths et al. 2016).

The first study, a randomized double-blind trial, was conducted by Ross and colleagues and placed 28 participants to receive Psilocybin or placebo (niacin), for one single session each.

Depression was evaluated using multiple depression scales (more details in supplementary table 1), and  they concluded that there were significant differences between the experimental and control groups. Psilocybin produced immediate and enduring anti-depressant response rates, as well as significant anti-depressant remission rates. For example, 7 weeks after dose 1, 83% of participants in the psilocybin first group (vs. 14% in the niacin first group) met criteria for anti-depressant response (with the BDI), (Ross et al. 2016).

The 6,5-months follow-up (after both groups received psilocybin) also showed great results, anti-depressant response rates were approximately 60–80% (Ross et all. 2016).

The second study, conducted by Griffiths and colleagues, which was also a randomized double-blind trial, enrolled 51 adults and compared a High dose of psilocybin (30mg/70kg and 20mg/70kg) to a low dose of the same drug, considered as inactive (3 and 1 mg/70kg).

The investigators also used multiple scales to assess the efficiency of psilocybin on depression such as GRID-HAM-D17 (details in supplementary table 1).

Results also assessed that Psilocybin was efficient to treat depression.

They found out that rates of clinically significant response and symptom remission of depression (GRID-HAMD-17) showed large effects of psilocybin that were sustained at 6 months.

For instance, 5 weeks after Session 1, 92% of participants in the High-Dose-1st Group showed a clinically significant response, compared with a 32% response rate in the Low-Dose-1st Group, at 6 months 79% of those in the High-Dose-1st Group continued to show a clinically significant response. (Griffiths et al. 2016) the overall rate of symptom remissions at 6 months for all participants was 65%, which is also significant.

## Meta analysis of the results :

We performed a Peto odds ratio with the extracted results at 5 to 7 weeks post dose 1 (depending the study) in order to examine response rates to the treatment but also remission rates for both studies (the results are exposed in table 4 and figure 2 and 3), we also combined these results, putting together the results obtained at the HADS-D scale from the study conducted by Ross et al. and at the GRID-HAM-D scale from the study conducted by Griffiths and colleagues, to evaluate psilocybin efficiency for depression across the studies.

**Table 4. Peto odds ratio results for depression 5 to 7 weeks after dose 1**

|  | Depression appreciation tool | OR (peto odds ratio) | CI (confiance interval) | V (Weight) |
|---|---|---|---|---|
| Study 1 : Ross et al. | BDI response rate | 0.069 | [0.057; 0.084] | V =1.813 |
|  | HADS-D response rate | 0.311 | [0.257; 0.375] | V=1.797 |
|  | BDI-Remission | 0.096 | [0.080; 0.117] | V=1.797 |
|  | HADS-D Remission | 0.099 | [0.082; 0.119] | V=1.815 |
| Study 2 : Griffiths et al. | GRID-HAM-D response rate | 0.099 | [0.086; 0.114] | V=3.029 |
|  | GRID-HAM-D Remission | 0.196 | [0.170; 0.226] | V=3.130 |
|  |  |  |  |  |
| Combined results. | Total Depression response rate | 0.149 | [0.133; 0.167] | V=4.792 |
|  | Total Remission for depression | 0.152 | [0.136; 0.169] | V=4.923 |

## Response

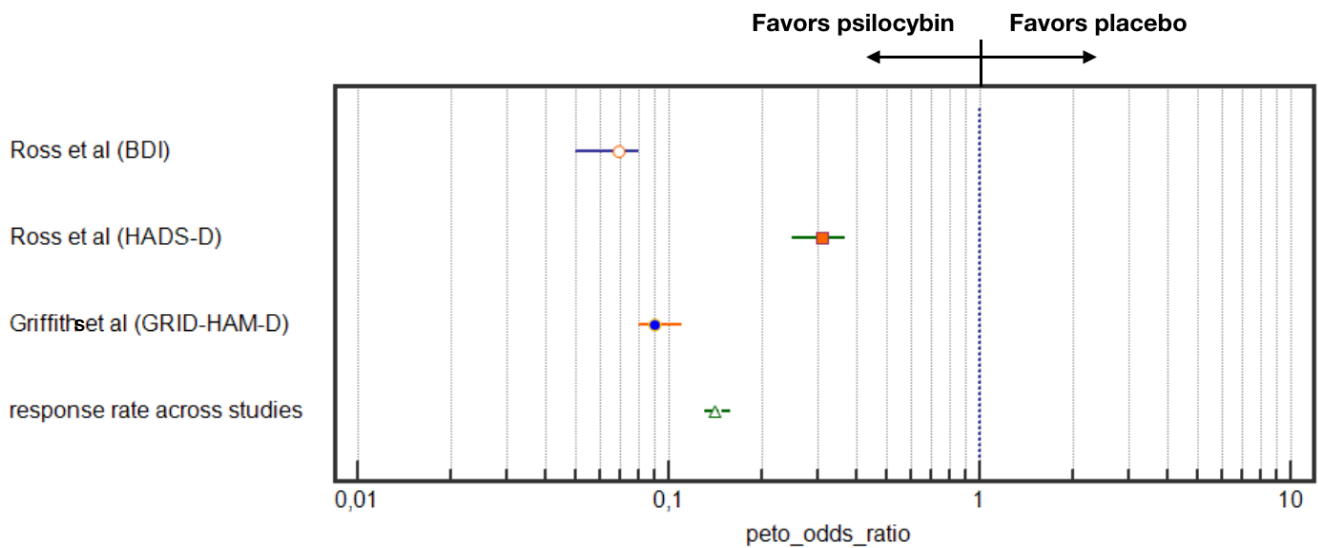

**Fig 2. Peto odds ratio for depression response rate**

Peto Odds performed on both studies showed an obvious superiority of psilocybin over placebo for Response rates to the treatment for depression.

## Remission

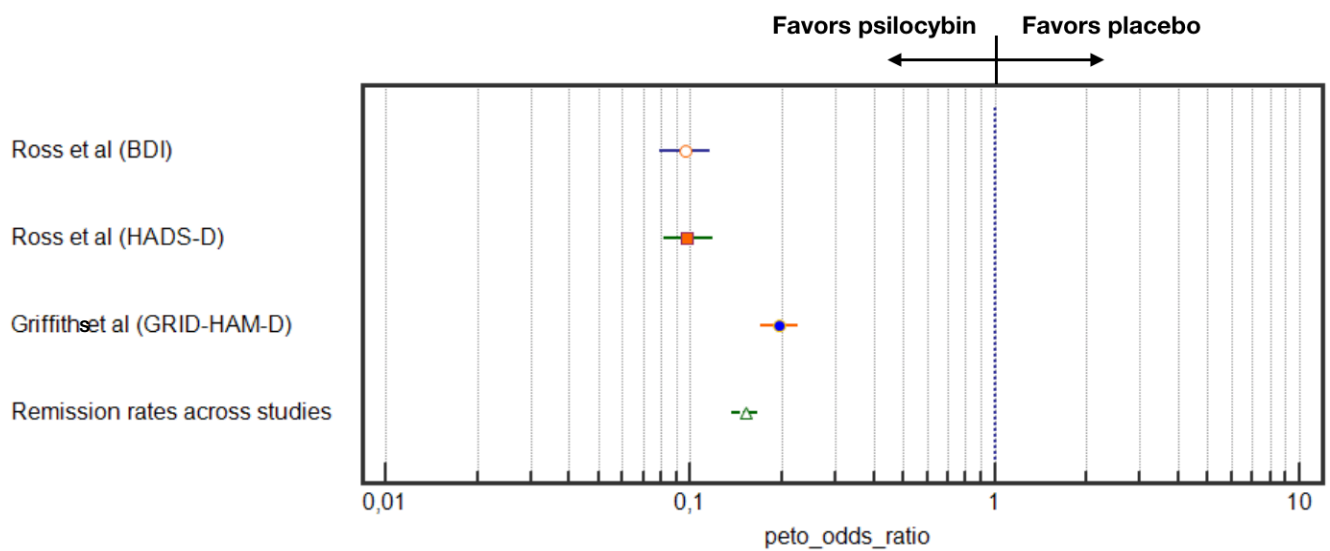

**Fig 3. Peto odds ratio for Remission rates to depression**

Peto Odds performed on both studies showed also a significant superiority of psilocybin over placebo for Remission to depression following the treatment.
Combined results assesses that superiority from psilocybin over placebo.

## Anxiety

The two quantitative studies also evaluated psilocybin as a treatment for cancer related anxiety (Ross et al. 2016, Griffiths et al. 2016).

The first study, a randomized double-blind trial, supervised by Ross and colleagues, placed 28 participants to receive Psilocybin or placebo (niacin), for one single session each.
Anxiety was evaluated using multiple Anxiety evaluation scales (more details in supplementary table 1), and concluded that the psilocybin group (compared to the active control) demonstrated immediate, substantial, and sustained (up to 7 weeks post-dosing) clinical benefits in terms of reduction of anxiety.
These reductions remained significant at approximately 8 months, post-psilocybin dosing. (Ross et al. 2016).

The second study, conducted by Griffiths and colleagues,  was also a randomized, double-blind trial, enrolled 51 adults and compared a High dose of psilocybin (30mg/70kg and 20mg/70kg) to a low dose of the same drug, considered as inactive (3 and 1 mg/70kg). (Griffiths et al. 2016).
The investigators also used multiple scales to assess the efficiency of psilocybin on Anxiety such as GRID-HAM A (more details in supplementary table 1).
Rates of clinically significant response and anxiety symptoms remission showed large effects of psilocybin that were sustained at 6 months. (Griffiths et al. 2016).

## Meta analysis of the results :

### Table 5. Peto odds ratio results for Anxiety 5 to 7 weeks after dose 1

| | Anxiety appreciation tool | OR (peto odds ratio) | CI (confiance interval) | V (weight) |
|---|---|---|---|---|
| Study 1 :Ross et al. | HADS-A response rate | 0.159 | [0.131;0.192] | V=1.672 |
| Study 2 : Griffiths et al. | GRID-HAM-A response rate | 0.141 | [0.122;0.162] | V=3.249 |
| | GRID-HAM-A Remission | 0.180 | [0.156;0.207] | V=2.972 |
| | | | | |
| **Combined results** | Total Anxiety response rate | 0.148 | [0.132;0.165] | V=4.956 |

A peto odds ratio of the extracted results obtained at 5 or 7 weeks (depending on study follow up) post dose 1 was performed in order to examine response rates to the treatment but also remission rates for both studies (results are exposed in table 5 and figure 4), we also combined the results, putting together the results obtained for the HADS-A scale from the study conducted by Ross and colleagues and for the GRID-HAM-A scale from the study conducted by Griffiths and colleagues to evaluate psilocybin efficiency on anxiety across the studies.

## Response

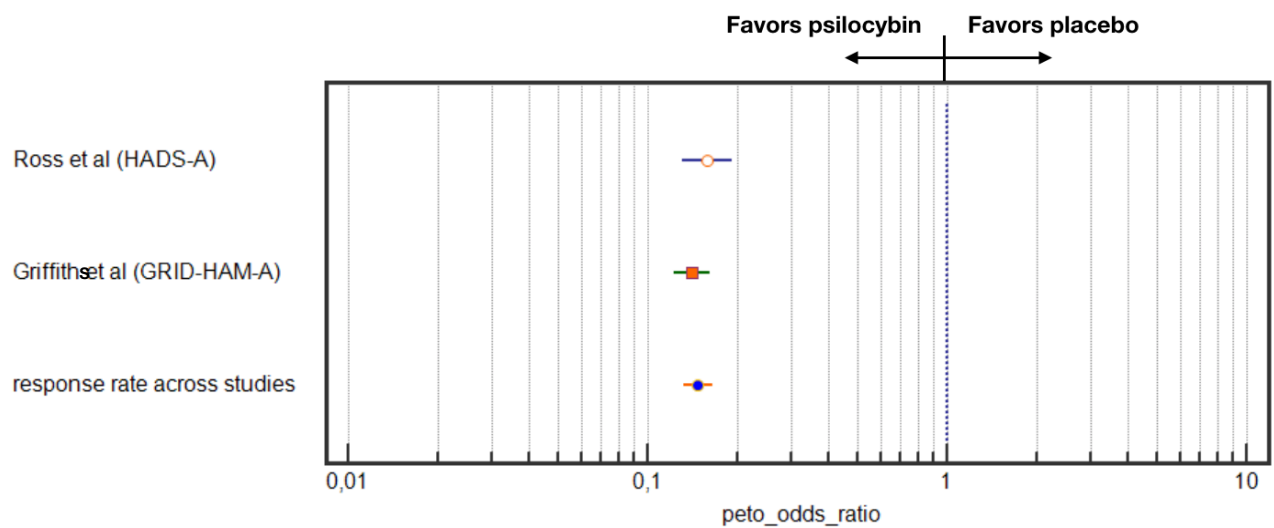

**Fig 4.** Peto odds ratio for anxiety response rate

Peto Odds performed on both studies showed a significant superiority of psilocybin over placebo for Response rates to the treatment for anxiety.

## Remission

As data for anxiety remission could not been obtained for the study conducted by Ross and Colleagues, we were not able to perform forest plot of the Peto odds ratio across the two studies.

It is important to note that the study conducted by Griffiths and colleagues concluded on a superiority of psilocybin over placebo for remission rates too.

## Psychiatric safety of use

The study conducted by Ross and Colleague found out that there were no serious Adverse events, either medical or psychiatric, in the trial that were attributed to either psilocybin or niacin. No pharmacological interventions (as benzodiazepines or anti-psychotics) were needed during dosing sessions, no participants abused or became addicted to psilocybin, there were no cases of prolonged psychosis or hallucinogen persisting perceptual disorder (HPPD) and no participants required psychiatric hospitalization.

They also noted that the most common psychiatric adverse events were transient anxiety (17%) and transient psychotic-like symptoms (7%: one case of transient paranoid ideation and one case of transient thought disorder), attributable to psilocybin and are all known adverse events of psilocybin, were transient, tolerable and consistent with previous research (Ross et al. 2016).

The second study, conducted by Griffiths and colleagues, determined that transient episodes of psychological distress during psilocybin sessions were more common after the high dose than the low dose.

They described occurrence of Psychological discomfort in 32% of participants in the high-dose session and 12% in the low-dose session. An episode of anxiety occurred in 26% of participants in the high-dose session and 15% in the low-dose session. One participant had a transient episode of paranoid ideation (in the high dose session) . There were no cases of hallucinogen persisting perception disorder or prolonged psychosis (Griffiths et al. 2016).

According to these results, psilocybin seems to be clinically safe for administration for cancer patients, in a supervised environment, on a psychiatric point of view.

## Medical safety

The first study, conducted by Ross and colleague, concluded that there were no serious adverse cardiac events.

They noted that the medical AEs (non-clinically significant elevations in blood pressure and Hearth rate, headaches, nausea) attributable to psilocybin were all known AEs of psilocybin, were transient and tolerable (Ross et al. 2016).

Later, the study supervised by Griffiths assumed that no serious adverse events attributed to psilocybin administration occurred.

A number of adverse events occurred during psilocybin sessions, none of which were deemed to be serious. All of these adverse events had resolved fully by the end of the sessions.

They noted that there were transient moderate increases in systolic and/or diastolic blood pressure after psilocybin. An episode of elevated systolic blood pressure occurred in 34% of participants in the high-dose session and 17% of participants in the low-dose session. An episode of elevated diastolic blood pressure occurred in 13% of participants in the high-dose session and 2% of participants in the low dose session, None of these episodes needed medical intervention.

Nausea or vomiting occurred in 15% of participants in the high-dose session and none in the low-dose session.

An episode of physical discomfort occurred in 21% of participants in the high-dose session and 8% in the low-dose session. (Griffiths et al. 2016)

According to these results we can assess that psilocybin seems to be safe to administer to cancer patients, in a supervised environment, on a medical point of view.

## Psychological mechanisms

The qualitative studies that were analyzed in this review focused on getting a better understanding of the psychological mechanisms, involved in a psilocybin therapy for depressive and anxious patients with cancer, and in depressive and anxious patients more globally.

The study conducted by Belser's team deducted that the psilocybin experience may be conceived as relationally embedded : indeed, the most common thematics evoked by participants were : forgiveness of others, loved ones as spirit guides, the importance of narrating one's experience with loved ones, and improved relationships posttreatment.

Nearly all participants reported having an experience of love and joy, though these exalted experiences arose primarily through a specific human relationship with another person, whether it was a parent, child, or partner (Belser et al. 2017).

Moreover, the findings of this study also suggest that embodiment, was a critical feature of participant experiences. (Belser et al. 2017)

Beside that, as slightly more than a half of participants experienced fear or anxiety during the sessions, they found out that transient periods of distress, even intense experiences of fear and panic, can be well tolerated by study participants in a supportive therapeutic milieu.

Both participants and the research team came to understand these difficult experiences as part of a necessary and ultimately beneficial process, they said.

These results also indicates that psilocybin occasioned a vast range of emotions, suggesting that participants may emerge from this treatment having experienced profound emotional depths and expanded affective boundaries (Belser et al. 2017).

Most of participants have experienced complexe closed eyes visuals, it seems that those may have served as principle organizing motifs of subjective experience with multifold vectors (e.g., audiovisual, relational, autobiographical, spiritual, epistemological, ontological). (Belser et al. 2017)

Finally, Belser and colleagues reported that participants described lasting impacts to their quality of life, life priorities, and their sense of identity, participants described feeling "reborn," more confident, more connected, and more alive. They also described a feeling of empowerment and being "unstuck," with resulting healthier behaviors. (Belser et al. 2017).

The second study, supervised by Swift and colleagues, revealed that experiential and immersive quality of the psilocybin therapy may help explain the immediate and enduring positive changes in participant's lives after a single session.

They noted that the sessions brought new perspectives to participants regarding their cancer and the way they experiences it, reporting that participants were able to more clearly consider the reality of cancer and death in their lives, and consistently reported no longer being preoccupied or overwhelmed by fear.

Swift and colleagues assessed that the mechanism behind the psilocybin sessions may not be apprehended like classical antidepressant or anxiolytic drugs, the results showed that following their psilocybin session, participants did not merely return to the level of functioning that existed prior to their cancer diagnosis, but rather nurtured a heightened sense of meaning and perspective in their lives. (Swift et al. 2017)

These studies suggests that Psylocybin can bring truly immersive and meaningful mystical experiences, sometimes leading participants to face their fears and change perspectives on something that has caused their psychological distress, making them able again to interact with others and with life in general.

This way, people learn how to live with their new condition, to accept it and eventually to use it to sublime and give a new sense to their life.

## Neurobiological mechanisms

The study conducted by Ross and colleagues, considered the existing literature to explore neurobiological mechanisms involved in the action of psilocybin :

Those are apparently linked to the 5HT2A binding effects of Psilocybin (as well as others psychedelics (serotoninergic hallucinogens), and more precisely, 5HT2A downregulation may explain some of the rapid and sustained anxiolytic effects of psilocybin.

Indeed, they reported that cortical 5HT2A receptor expression is increased in postmortem samples of patients with depression who display suicidality.

This study also suggest that the glutamate system may explain some of the anti-depressant effects of psilocybin : In rodents, serotoninergic psychedelics enhance cortical glutamatergic transmission, especially in the medial PFC, and increase activation of cortical α-amino-3-hydroxy-5-methyl-4-isoxazolepropionic acid (AMPA) receptors (Aghajanian and Marek, 1997) as well as BNDF.

In parallel, cortical AMPA activation is known to stimulate the expression of cortical BDNF (associated with neuronal growth, differentiation and synaptogenesis) (Hsu et al. 2015) and decreased cortical BDNF is associated with major depression in humans (Duman. 2004). Also, cortical BDNF normalizes with anti-depressant treatment (Shimizu et al. 2003).

Moreover, neuroimaging research with psilocybin is beginning to suggest potential anti-depressant mechanisms of action at the level of brain structure activity and network connectivity, they said.

Task-free functional magnetic resonance imaging research in normal volunteers under the influence of psilocybin has demonstrated decreased activity in the medial PFC and decreased connectivity within the default mode network (DMN) (Carhart-Harris et al. 2012, 2014), beside that, depressive symptoms have been associated with increased activity in the medial PFC (Farb et al. 2011) and normalization of medial PFC activity has been demonstrated with anti-depressant treatment (Holtzheimer and Mayberg, 2011), Patients with major depression (compared to controls) also demonstrated increased DMN connectivity (Berman et al. 2011), they said.

Complexe neurobiological mechanisms are involved in psilocybin action and are yet to be established. This may help to better understand brain connectivity and depression or anxiety mechanisms. The Glutamatergic system as well as the activity modulation in the DMN seems to represent some promising key mechanisms.

## Discussion

This systematic review and meta-analysis revealed the beginning of a promising new approach about how to treat cancer related distress, and maybe even more, anxiety and depression in a more general way.

Since this review focuses on a really innovative approach, it is the first to be realized about the subject.

For the same reason, the number of studies included was relatively limited : indeed, we were able to include only 4 studies comprising 105 participants, synthesizing data about 6 different themes. Psilocybin seemed to have a real potential for mental healthcare.

Regarding to the results, this review might open minds about psilocybin and more widely, psychedelics (serotoninergic hallucinogens), that inherited a sulfurous reputation since the 70's and help to bring interest about these substance, that may completely renew the way we approach mental illness and it's treatment, by it's very specific action mechanism, both on a psychological and neurobiological point of view. These kind of therapies may improve quality of life of many distressed people, as cancer patients.

This review also underline the ambitious and passionate work Heffter institute achieved and continue to realize, about psychedelics for mental healthcare.

## Main findings : potential use for depression and anxiety in cancer patients.

The results exposed in this review showed that psilocybin appear to be effective as a treatment for depression and anxiety, obtaining significant results on multiple and recognized depression and anxiety measuring tools, performed meta analysis (which consisted in a Peto odds ratio) also concluded a significant superiority of a single psilocybin session over placebo for both depression and anxiety, as no classical anti-depressant ever reached this level of proof for treating cancer related distress. (Ostuzzi et al. 2015)

The remission rates were also significant for anxiety and depression on different scales such as BDI, HADS-D and GRID-HAM-D and GRID-HAM A.

Moreover, for one single session, the results sustained at least 6-months post session for 60 to 80 % of patients overall.

Regarding these results Psilocybin seems to be surprisingly efficient for treating cancer related depression and anxiety.

Numerous recents studies also showed great results for others use, as : addiction caring, PTSD, depression and anxiety in non cancerous patient and a great part of it were conducted by the Heffter institute.

## Safety of use

Globally, no serious adverse events have been reported across the studies, on a psychiatric point of view as well as a medical one, these results seems to indicates that psilocybin is safe to use in a medically supervised environment, even for sensible patients like cancer patients, the results were consistent to prior research on healthy volunteers and cancer patients, but the relatively limited size of the sample included in this review, underlines the fact that it has to be confirmed on larger populations. It is to know that very recents studies focusing on the safety with larger samples confirmed these results, that the Global Drug survey of 2017 designed it as the safest of any recreative drug (Winstock et al. 2017) and that the FDA recently approved psilocybin for a wider, phase 2b trial including on 216 participants across North America and Europe (Compass pathways. 2018), which is really encouraging.

## Psychological and Neurobiological mechanism

This review can't conclude on a precise psychological or neurobiological mechanism involved in the Psilocybin mode of action, but the results demonstrated complexe and previously unthanked mechanisms, both on a psychological and neurobiological point of view.

Indeed, learning more about those, could bring a better knowing of general mental health, mechanisms of certain mental diseases like depression, anxiety, distress relating to end of life, and more generally, overall consciousness function.

This way, it could also help to enhance psychological therapies and the way we care cancer patients and mental illnesses in general.

Thanks to psilocybin, patients don't return to the state they have been prior their cancer, but learn to cope with it and to change their perspective toward cancer, accepting it and subliming it, to get new aims in life as well as a need of reconnection to others, their family, etc.

Psilocybin seems to make patient face their fear and the reality of what they try to avoid, sometimes triggering challenging experiences, but experiences that serve as a support of their personnal sublimation or accomplishment. Those moments seems to represent an occasion to mature and grow-up off their difficulties, according to participants.

On a neurobiological point of view, research on the subject seems to indicates that the anti depressive effect of psilocybin may be related to it's aptitude to provoke 5HT2A receptors down-regulation, enhancements of glutamatergic transmission in the medial PFC that trigger α-amino-3-hydroxy-5-methyl-4-isoxazolepropionic acid (AMPA) activation, which stimulate BDNF production (as a decrease is associated with depression in human). Neuroimaging reported a normalized medial PFC activity (hyperactive region in depressed patients) and a decreased DMN (default mode network) activity, while it was increased in depressed patients.

Psilocybin seems to share some neurobiological aspects with classical antidepressants, but also many differences both on a psychological and neurobiological point of view. These mechanisms are still to investigate and not well known or understood, but these elements are encouraging and may represent some new key mechanisms to explore and the constant development in neurobiology and neuroimaging as well as the growing knowledge in psychology, should bring some further evidences, ideas and reflection about the topic.

## Strengths and Limitations

This comprehensive systematic review includes a global look around Psilocybin therapy, focusing on both quantitative aspects (with a serie of exploratory meta-analysis about the efficiency of this serotoninergic psychedelic on cancer related distress (that were examined through the scope of multiple depression and anxiety appreciation tools)), and qualitative aspects (with the narrative analysis of qualitative studies about the psychological mechanisms underlaying those therapeutical effects).

This review also sums-up and bring the lights on the significant effort and implications put by Heffter institute in research about Psilocybin based therapies and more widely, psychedelic therapies for mental healthcare. Those topics have, for sure, a promising future.

Notwithstanding this, there are several limitations and results should be interpreted with some caution.

First of all, this review includes a relatively limited amount of studies, with a relatively limited amount of participants, in this regard, the findings should be reevaluated when more studies including wider populations will be released.

Most of the patients were Caucasian and highly educated, which may influence the experiences participants lived during the session and decrease the overall generalizability.

The strength of the findings exposed in this review result on the high quality of the studies and the apparent low risk of bias using tools like Cochrane Risk of Bias Tool for Randomized Controlled Trials, Jadad scale and Critical Appraisal Skills Program for qualitative studies.

Another limitation is that the meta analysis regrouped results that were obtained through different depression and anxiety scales, the appreciation criteria that constitute those scales are not exactly the same. In other hands, the choice of using a Peto odds ratio method was judicious because the participants groups were balanced in size, but since the difference from placebo revealed to be very significant, it may have exaggerated the results slightly.

Also, this study does not really take in account the psychotherapies sessions occurring before, during and after the sessions, that are an essential component of the patient care.

Regarding the supposed psychological action mechanism of psilocybin, it is very complementary with those type of therapies (psychotherapy), certainly participating to the really encouraging results.

The crossover between psilocybin and placebo groups after 5 weeks may also influence the results at the 6 months follow-up, specifically for the low dose session.

# Conclusion and future implications

Through this review, we discovered that Psilocybin seemed to be safe and really effective for treating cancer related depression and anxiety, even with one single dosing session.

Moreover, it has uncovered that the mechanisms responsible of the anti-depressant and anxiety effect are obviously very particular and seems to change people's perceptions, leading them to acceptance and to integrate their struggles into something « useful » for their lives, instead of negatively searching to get rid of it, to avoid it, Fearing it.

The immersive and sometimes challenging experiences appear to serve as a vector of integration to the knowledge participants received during the session, which is often described as « beyond words ». (Belser et al. 2017)

On the other hand, no persisting or serious psychiatric adverse event has been signaled among the studies.

Apparently, challenging experiences should be appreciated as a potentially useful and sometime necessary part of the experience. This underlines the importance of a supervised environment and the integration of psychotherapy sessions to optimize the results.

Beyond the necessity to verify the reproducibility of these results on larger populations, the question of it's efficiency for depression and anxiety in healthy and non hospitalized participants, and the question of how to organize supervised sessions and follow-up with these type of ambulatory participants may be reached.

Microdosing with repeated dosing sessions, appear to bring promising results in some recents studies, avoiding the challenging side of the experience, which might bring some autonomy to the patients, for this kind of therapy.

In other hand neurobiological mechanisms are still to precise and certainly hide some revelations about mechanisms involved in depression, consciousness and spirituality.

# BAHI Camile, 2018

## Supplementary table 1 :

| | Author | Year | Country | Percentage of Women | Date of the sample | Type of cancer | Stage | Follow-up | Prior Depression/ condition | Studied outcomes | Measurement tool | Serious adverse event/ psychological adverse event | Adverse events | Study setup | Dosage/sessions | Dosages | Frequency Rating scale (for quantitative study) |
|---|---|---|---|---|---|---|---|---|---|---|---|---|---|---|---|---|---|
| **Study 1** | Stephen Ross, Anthony Bossis, Jeffrey Guss, Gabrielle Agin-Liebes, Tara Malone, Barry Cohen, Sarah E. Mennenga, Alexander Belser, Krystallia Kalliontzi, James Babb, Zhe Su, Patricia Corby, L. Schmidt | 2016 | USA | 62 % | | 28 - Breast 11% - Reproductive 24% - Digestive cancers 17% - Lymphoma/leukemia 14% - Other types 10% | Stage IV 34%; Stage III 29%; Stage II 5%; Stage I 5%; Other 5% | | BPDIB | Cardiovascular measures and mental ratings; - Hospital Anxiety and Depression (HAD); - quality of life | None | | | | | |
| **Study 2** | Roland R. Griffiths, Matthew W. Johnson, Michael A. Carducci, Annie Umbricht, William A. Richards, Brian D. Richards, Mary P. Cosimano, Margaret A. Klinedinst | 2016 | USA | 49 % | | 51 - breast; upper; gastrointestinal; other | 51 | | | | | None | | | | | |
| **Study 3** | Alexander B. Belser, Gabrielle Agin-Liebes, T. Cody Swift, Sara Terrana, Neşe Devenot, Harris L. Friedman, Jeffrey Guss, Anthony Bossis, Stephen Ross | 2017 | USA | 46 % | | 13 - Breast - Lymphoma - Other - Ovarian | NI | | | | | None | | | | | |
| **Study 4** | Thomas C. Swift, Alexander B. Belser, Gabrielle Agin-Liebes, Neşe Devenot, Harris L. Friedman, Jeffrey Guss, Anthony Bossis, Stephen Ross | 2017 | USA | 62 % | | 13 - Breast - Lymphoma - Other - Ovarian | NI | | | | | None | | | | | |

# Acknowledgements

The author would like to thank Pr. Haffen (MD, PhD) as well as Dr. Nicolier (PhD) for evaluating his work and giving advices about this systematic review as part of a university presentation.

# Declaration of the conflicting interests

# Abstract

**Background :**

Depression and anxiety are common in patients with cancer, classical antidepressant and anxiolytics has no proven efficacy on this type of distress compared to placebo. A Psilocybin (serotoninergic hallucinogen) based therapy appear to give promising results among some recent studies.

**Aims :**

To examine if a psilocybin based therapy could be considered for patients with cancer related depression and/or anxiety and it's safety.

To sum Heffter's institute work, as the main research institute working on this topic.

**Method :**

Following PRISMA a guidelines, a systematic review was conducted, for quantitative and qualitative studies about psilocybin for treating cancer related depression and anxiety. Pubmed and the Heffter institute database has been reached for this purpose, separating studies in types : qualitative or quantitative. We studied the effects on cancer related depression and anxiety separately and investigated the psychological and neurobiological mechanisms.

**Results :**

The 4 studies included a total of 105 randomized participants, meta analysis on depression and anxiety with pooled Peto odds ratio showed a significant superiority of psilocyfbin over placebo. The substance appeared to be safe for this type of patients. Surprising psychological mechanisms hypothesis have been found out.

**Conclusions :**

Psilocybin appear to be potentially useful as treatment for cancer related depression and anxiety. Future research should verify these findings on wider populations and eventually seek a way to apply this therapy to non hospitalized patient.

**Keywords** :

psilocybin, depression, anxiety, review, meta-analysis

**Contexte** :

La dépression et l'anxiété sont fréquentes chez les patients cancéreux, les antidépresseurs classiques et les anxiolytiques n'ont pas prouvé leur efficacité sur ce type de détresse par rapport au placebo. Une thérapie basée sur la psilocybine (hallucinogène sérotoninergique) semble donner des résultats prometteurs parmi certaines études récentes.

**Objectifs:**

examiner si un traitement à base de psilocybine pourrait être envisagé chez les patients souffrant de dépression et / ou d'anxiété liée au cancer et si il peut être utilisé en toute sécurité, Mais également de résumer le travail de l'institut Heffter, en tant que principal institut de recherche travaillant sur ce sujet.

**Méthode :**

En se référant aux lignes directrices PRISMA (Preferred Reporting Items for Systematic reviews and Meta-analyses), une étude systématique a été menée, pour les études quantitatives et qualitatives sur la psilocybine dans le traitement de la dépression et de l'anxiété liées au cancer. Pubmed et la base de données de l'institut Heffter ont été consultés à cette fin, en séparant les études par types : qualitative ou quantitative. Nous avons étudié les effets sur la dépression et sur l'anxiété liées au cancer séparément et ainsi que les mécanismes psychologiques et neurobiologiques.

**Résultats :**

Les 4 études on inclus un total de 105 participants randomisés, une méta-analyse sur la dépression et l'anxiété grace a un Peto odds ratio combiné a montré une supériorité significative de la psilocybine par rapport au placebo. La substance semble également sans danger pour ce type de patients. Des hypothèses de mécanismes psychologiques inattendues ont été découvertes.

**Conclusions:**

La psilocybine semble potentiellement utile comme traitement contre la dépression et l'anxiété liées au cancer. Les futures recherches devraient vérifier ces résultats sur des populations plus larges et éventuellement chercher un moyen d'appliquer cette thérapie à des patients non hospitalisés.

**Mots clés :**

psilocybine, dépression, anxiété, revue de literature, méta-analyse.